\documentclass[journal]{IEEEtran}
\ifCLASSINFOpdf

\else

\fi
\usepackage[cmex10]{amsmath}
\usepackage{amssymb}
\usepackage{cite}
\usepackage{graphicx}
\usepackage{array,color}
\usepackage{multirow}
\usepackage{amsmath}
\usepackage{stfloats}
\usepackage{graphicx}
\usepackage{subfigure}
\usepackage{tabularx}
\usepackage{epsfig,epsf,color,balance,cite}
\usepackage{setspace}
\usepackage{bm}
\usepackage{textcomp}
\usepackage[linesnumbered, ruled]{algorithm2e}
\usepackage{subfigure}
\usepackage{caption}
\usepackage{graphicx}
\usepackage{setspace}
\SetKwRepeat{Do}{do}{while}%

\begin{document}

\title{Joint Transmit Power and Placement Optimization for URLLC-enabled UAV Relay Systems}
\author{Hong Ren,  Cunhua Pan, Kezhi Wang, Wei Xu, Maged Elkashlan, and Arumugam Nallanathan, \IEEEmembership{Fellow, IEEE}
\thanks{H. Ren, C. Pan,  M. Elkashlan and A. Nallanathan are with School of Electronic Engineering and Computer Science, Queen Mary University of London, London, E1 4NS, U.K. (Email:{ h.ren, c.pan, maged.elkashlan, a.nallanathan}@qmul.ac.uk). K. Wang is with Department of Computer and Information Sciences,
Northumbria University, UK. (e-mail: kezhi.wang@northumbria.ac.uk). W. Xu is with NCRL, Southeast University, Nanjing 210096, China (e-mail: wxu@seu.edu.cn). A. Nallanathan would like to thank the U.K. Engineering and the Physical Sciences Research Council under Grant EP/N029666/1. W. Xu would like to thank the NSFC under grants 61941115 and 61871109. (\emph{Corresponding author: Cunhua Pan, Kezhi Wang.})}
}

\maketitle

\begin{abstract}
This letter considers an unmanned aerial vehicle (UAV)-enabled relay communication system for delivering latency-critical messages with ultra-high reliability, where the relay is operating under amplifier-and-forward (AF) mode. We aim to jointly optimize the UAV location and power to minimize decoding error probability while guaranteeing the latency constraints. Both the free-space channel model and three-dimensional (3-D) channel model are considered. For the first model, we propose a low-complexity iterative algorithm to solve the problem, while globally optimal solution is derived for the case when the signal-to-noise ratio (SNR) is extremely high. For the second model, we also propose a low-complexity iterative algorithm to solve the problem. Simulation results confirm the performance advantages of our proposed algorithms.
\end{abstract}


\IEEEpeerreviewmaketitle
\vspace{-0.6cm}
\section{Introduction}

Recently, unmanned aerial vehicle (UAV) communication has received considerable research interests due to its flexible deployment and the dominance of line-of-sight links \cite{yongzengmaga,huizhao}. UAV can be deployed as a relay when there is no direct link between any two nodes. Specifically, Zeng \emph{et al.} \cite{yongzeng2016} first studied the trajectory and power allocation for UAV-relay systems. In \cite{yunfeichen}, the reliability  of the UAV relay was analyzed in terms of outage probability and bit error rate.
%

On the other hand, ultra-reliable and low-latency communications (URLLC) have been regarded as one of the three important use cases in 5G \cite{Shafijsac}. For URLLC, a transmitter usually sends a short packet such as command signals or measurement data to a receiver, in contrast to conventional human-to-human communication where long packet is normally transmitted. Hence, a direct result of the  Shannon's capacity based on the law of large numbers may not be applicable. In \cite{Polyanskiy2010IT}, Peter \emph{et al. } have derived the maximal coding rate for short-packet transmission, which is a complicated function of channel blocklength and SNR.

In this paper, we consider a two-dimensional UAV-enabled industrial automation scenario in Fig.~\ref{systemodel}, where a controller needs to send command messages to a distant robot that conducts an experiment in a multi-hazard area. For the safety of workers, shelters such as thick cement/metal walls are built between the robot and the controller. Hence, the channel gain between the controller and the robot is weak and negligible, and requires a UAV to fly above the shelter to assist the transmission between the controller and the robot.  In \cite{cpan2019}, we  studied the problem of jointly optimizing the blocklength and location for UAV-relay communication systems, where the decoding-and-forward (DF) protocol was considered. However, additional processing time is required for the DF mode, which may not be applicable to URLLC applications. Motivated by above, we jointly optimize power and location to minimize the decoding error probability, where the relay is operating under the AF mode without the signal processing delay. The decoding error probability under short blocklength is adopted. We first prove that the decoding error probability is a monotonically decreasing function the SNR. Then, two channel models are studied: free-space channel model and the 3-D channel model. For the first one, an iterative algorithm is proposed to obtain the suboptimal solution with low complexity, and the globally optimal solution is obtained in closed form when the SNR is extremely high. Simulation results show the performance advantages of our proposed algorithms.

\vspace{-0.2cm}
\section{System Model}\label{system}

\begin{figure}
\centering
\includegraphics[width=2.4in]{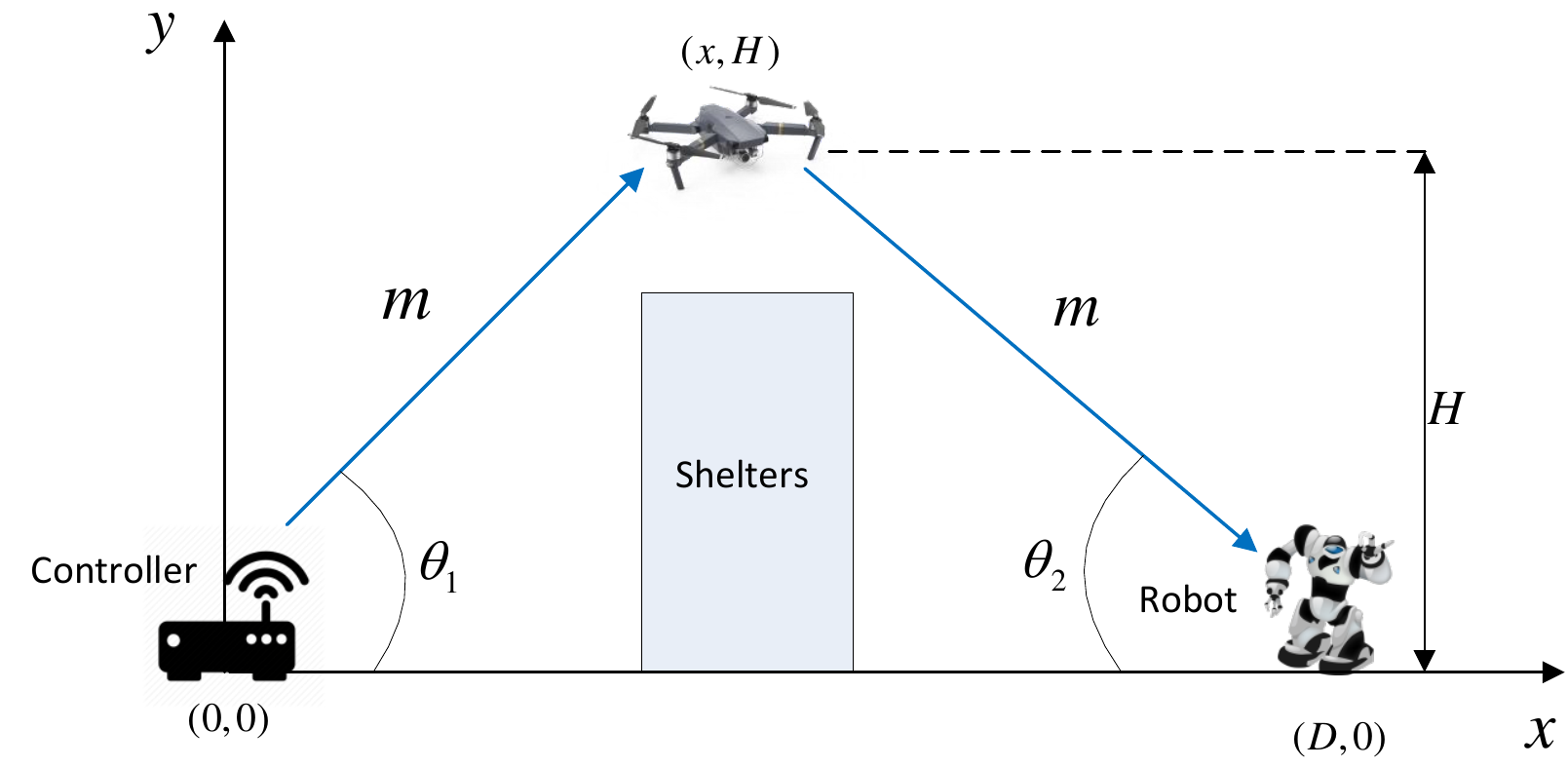}
\caption{UAV  relay system for delivering URLLC services.}
\vspace{-0.7cm}
\label{systemodel}
\end{figure}

As shown in Fig.~\ref{systemodel}, we consider a two-dimensional UAV-enabled industrial automation scenario, where the UAV hovers at a location $(x,H)$ above the horizontal line with height $H$. The locations of the controller and the robot are   $(0,0)$ and $(D,0)$. The packet size of the command information is $L$ bits, which should be completed within $T_{\rm{max}}$ seconds. Then, the overall blocklength is given by $M=BT_{\rm{max}}$ \cite{Durisi2016}, where $B$ is system bandwidth. The total transmission has two phases, i.e., the first one  corresponds to the transmission from the controller to the UAV, while the second one is the transmission from the UAV to the robot. We assume that the UAV adopts the AF protocol, which simply amplifies and forwards the received signals to the robot. Hence, the blocklength for these two phases should be equal, e.g., $m_1=m_2=m\triangleq  M/2$.  The transmit power of the controller and the UAV are respectively  $p_1$ and $p_2$.

The channel power gain from the controller to the UAV and from the UAV to the robot are denoted as $h_1$ and $h_2$, respectively. These channel gains depends on the height $H$ and horizontal distance $x$. In the first phase, the received signal at UAV is given by ${y_1} = \sqrt {{p_1}{h_1}} {x_1} + {n_1}$, where $x_1$ is the command signal transmitted by the controller with unit power, and $n_1$ is the received noise at the UAV that is normalized to unit. The amplification coefficient at the UAV is given by $G = \sqrt {{{{p_2}} \mathord{\left/
 {\vphantom {{{p_2}} {\left( {{p_1}{h_1} + 1} \right)}}} \right.
 \kern-\nulldelimiterspace} {\left( {{p_1}{h_1} + 1} \right)}}} $. In the second phase, the received signal at the robot is given by ${y_2} = \sqrt {{h_1}{h_2}{p_1}} G{x_1} + \sqrt {{h_2}} G{n_1} + {n_2}$, where $n_2$ is the noise power at the robot that is normalized to unit. Then, the received signal-to-noise ratio (SNR) at the robot is
 \begin{equation}\label{sdWDE}
   \gamma  = \frac{{{h_1}{h_2}{p_1}{p_2}}}{{{h_2}{p_2} + {h_1}{p_1} +1}}.
 \end{equation}

 In \cite{Polyanskiy2010IT}, the packet error probability of the AF relay system in short blocklength region can be approximately as:
 \begin{equation}\label{wearf}
{\varepsilon } = Q\left( {f\left( {{\gamma },{m},L} \right)} \right),
\end{equation}
where $f\left( {\gamma, m, L} \right) = \ln2\sqrt {\frac{m}{V(\gamma)}} \left( {{{\log }_2}(1 + \gamma) - \frac{L}{m}} \right)$, $V(\gamma)$ is the channel dispersion given by $V(\gamma)=1-(1+\gamma)^{-2}$ \cite{Polyanskiy2010IT}, and $Q\left( x \right)$ is the Gaussian $Q$-function.

In this paper, we aim to minimize ${\varepsilon }$ by optimizing the power allocation and the location of the UAV:
\begin{subequations}\label{initial-pro1}
\begin{align}
\mathop {\min }\limits_{\left\{ {{p_1},{p_2},x,H} \right\}} \;\;\;&  {{ \varepsilon }}\\
{\rm{s.t.}}\;\;\;& d_1\le x\le d_2,\label{djreitjgot}\\
& H_{\min}\le H\le H_{\max}, \label{fjroigtjio}\\
&p_1+p_2 \le  P_{T},\label{freocdsdi}\\
&p_1\ge 0, p_2\ge 0, \label{ofrrefetypo}\vspace{-0.1cm}
\end{align}
\end{subequations}
where constraint (\ref{djreitjgot}) and (\ref{fjroigtjio}) specifies the feasible flying region of the UAV, and $P_T$ is the total power limit.

Before solving Problem  (\ref{initial-pro1}), we first provide the following lemmas.

\emph{\textbf{Lemma 1}}: The packet decoding error probability $\varepsilon $ is a decreasing function of SNR $\gamma$.

\emph{\textbf{Proof}}: \upshape Please refer to  Appendix \ref{prooflemma1}. \hfill\rule{2.7mm}{2.7mm}

Then, Problem (\ref{initial-pro1}) can be equivalently formulated as
\begin{subequations}\label{inDAfefro1}
\begin{align}
\mathop {\max }\limits_{\left\{ {{p_1},{p_2},x,H} \right\}} \;\;\;&  \gamma \\
{\rm{s.t.}}\;\;\;& (\ref{djreitjgot}), (\ref{fjroigtjio}), (\ref{freocdsdi}), (\ref{ofrrefetypo}).
\end{align}
\end{subequations}

Then, we have the following lemma.

\emph{\textbf{Lemma 2}}: The total power constraint in (\ref{freocdsdi}) holds with equality at the optimal solution.

\emph{\textbf{Proof}}: \upshape This can be proved by using contradiction, the details of which are omitted due to limited space. \hfill\rule{2.7mm}{2.7mm}

It is difficult to obtain the globally optimal solution of  Problem (\ref{inDAfefro1}) because the power allocation are coupled with the location. In the following, we first consider the free-space channel model, and then we consider the more practical 3-D channel model.

\section{Free-Space Channel Model}\label{proformu}

In this section, we  assume the channel is dominated by line-of-sight (LOS) component, and  consider the free space channel model, i.e.,
\begin{equation}\label{edeaftg}
 {h_1} = \frac{{{\beta _1}}}{{{H^2} + {x^2}}},\ {h_2} = \frac{{{\beta _2}}}{{{H^2} + {{(D - x)}^2}}},\vspace{-0.1cm}
\end{equation}
where $\beta _1$ and $\beta _2$ are channel power gains at a reference distance of $d=1$ meter for the two links, respectively. In this case, we fix the height $H$, and optimize the power and horizontal distance $x$. Then, Problem (\ref{inDAfefro1}) becomes
\begin{subequations}\label{nfhoij}
\begin{align}
\mathop {\max }\limits_{\left\{ {{p_1},{p_2},x} \right\}} \;\;\;&  \gamma \\
{\rm{s.t.}}\;\;\;& (\ref{djreitjgot}),  (\ref{freocdsdi}), (\ref{ofrrefetypo}).
\end{align}
\end{subequations}

In the following, we first consider the general case and solve the problem by using the block coordinate decent (BCD) method. Then, we consider the special case when the SNR is extremely high, where the globally optimal solution can be obtained.
\subsection{General Case-BCD method}
In the following, we decouple Problem (\ref{nfhoij}) into two subproblems, i.e., optimize power allocation with fixed $x$ and
vice versa. Then, iteratively solve these two subproblems until convergence.
\subsubsection{Power Allocation with Fixed $x$}\label{joijiy}
Given $x$, Problem (\ref{inDAfefro1}) can be transformed to the following subproblem:
\begin{subequations}\label{defrgtwss}
\begin{align}
\mathop {\max }\limits_{\left\{ {{p_1},{p_2}} \right\}} \;\;\;&  \gamma\\
{\rm{s.t.}}\;\;\;& (\ref{freocdsdi}), (\ref{ofrrefetypo}).\vspace{-0.1cm}
\end{align}
\end{subequations}
By substituting $p_2=P_T-p_1$ into the expression of $\gamma$ in (\ref{sdWDE}) and performing some  manipulations, $\gamma$ can be rewritten as:
\begin{equation*}
\gamma  \!= \! - \!\frac{{{h_1}{h_2}}}{{{A^2}}}\left(\! {A{p_1} + B}\! \right)\! -\! \frac{{\frac{{{h_1}{h_2}B}}{A}\left( {\frac{B}{A}\! +\! {P_T}} \right)}}{{A{p_1} \!+\! B}}\! + \! \frac{{{h_1}{h_2}}}{A}\left(\! {\frac{{2B}}{A} \!+ \!{P_T}} \!\right),
\end{equation*}
where $A=h_1-h_2$ and $B=P_Th_2+1$. The second order derivative of $\gamma$ w.r.t. $p_1$ is calculated as
\[\frac{{{\partial ^2}\gamma }}{{\partial p_1^2}} =  - \frac{{2{h_1}{h_2}B\left( {B + A{P_T}} \right)}}{{{{\left( {A{p_1} + B} \right)}^3}}}\]
which can be checked to be negative. Hence, $\gamma $ is a concave function and the optimal solution of Problem (\ref{defrgtwss}) can be derived as follows:
\begin{equation}\label{dfregh}
 p_1^* = \frac{{\sqrt {B\left( {B + A{P_T}} \right)}  - B}}{A},\ p_2^* = {P_T} - p_1^*.
\end{equation}

\vspace{-0.2cm}\subsubsection{Location Optimization with Fixed $p_1$ and $p_2$}
By substituting the expressions of $h_1$ and $h_2$ in (\ref{edeaftg}) into the expressions of $\gamma$, Problem (\ref{inDAfefro1}) is equivalent to
\begin{subequations}\label{joijouopi}
\begin{align}
\mathop {\min }\limits_{x} \;\;\;&  {p_2}{\beta _2}{D_1(x)} + {p_1}{\beta _1}{D_2(x)} + {D_1(x)}{D_2(x)}\\
{\rm{s.t.}}\;\;\;& d_1\le x\le d_2,\vspace{-0.1cm}
\end{align}
\end{subequations}
where $D_1(x)=H^2+x^2$ and $D_2(x)=H^2+(D-x)^2$. Obviously, the objective function (OF) in Problem (\ref{joijouopi}) is a continuous function, and the globally optimal solution of Problem (\ref{joijouopi}) is among the locally optimal solutions and boundary points. By setting the first derivative of OF w.r.t. $x$ to zero, we have
\begin{equation}\label{johpipkww}
a{x^3} + b{x^2} + cx + d = 0,
\end{equation}
where $a = 4$, $b =  - 6D$, $d =  - 2\left( {D{H^2} + D{p_1}{\beta _1}} \right)$, and $c = 2\left( {{D^2} + 2{H^2} + {\beta _1}{p_1} + {\beta _2}{p_2}} \right)$.
Dividing equation (\ref{johpipkww}) by $a$ and substituting $x=t-b/3a$, we have
\begin{equation}\label{swdeartrsg}
{t^3} + \rho t + \kappa  = 0,
\end{equation}
where $\rho  = \frac{{3ac - {b^2}}}{{3{a^2}}}$ and $\kappa  = \frac{{2{b^3} - 9abc + 27{a^2}d}}{{27{a^3}}}$.

Equality (\ref{swdeartrsg}) is a cubic equation. The equation may have only one real solution or three solutions, which depends on the conditions. Specifically, if $4{\rho ^3} + 27{\kappa ^2} > 0$ and $\rho  < 0$, there is only one real solution, given by
\begin{equation}\label{jojytjhy}
 {t_0} =  - 2\frac{{\left| \kappa  \right|}}{\kappa }\sqrt { - \frac{\rho }{3}} \cosh \left( {\frac{1}{3}{\rm{arcosh}}\left( {\frac{{ - 3\left| \kappa  \right|}}{{2\rho }}\sqrt {\frac{{ - 3}}{\rho }} } \right)} \right),
\end{equation}
if $\rho  > 0$, there is only one real solution, given by
\begin{equation}\label{kkhlol}
{t_0} =  - 2\sqrt {\frac{\rho }{3}} \sinh \left( {\frac{1}{3}{\rm{arsinh}}\left( {\frac{{3\kappa }}{{2\rho }}\sqrt {\frac{3}{\rho }} } \right)} \right),
\end{equation}
otherwise, there are three real solutions given by
\begin{equation}\label{koiylyol}
\!\!\!{t_k} \!=\! 2\sqrt {\! - \frac{\rho }{3}}\! \cos \left(\! {\frac{1}{3}{\rm{arcos}}\left(\! {\frac{{3\kappa }}{{2\rho }}\sqrt {\frac{{ - 3}}{\rho }} }\! \right) \!-\! \frac{{2\pi k}}{3}} \!\right),k = 0,1,2.
\end{equation}

Once obtaining the real solution of (\ref{swdeartrsg}), set $x_0=t_0-b/3a$ for only one real solution, and $x_k=t_k-b/3a, k=0,1,2$ for three different real solutions. For the one real solution case, if $x_0$ is in the range of $[d_1,d_2]$, choose one from the set $\{d_1,d_2,x_0\}$ with the minimum OF of Problem (\ref{joijouopi}), otherwise, choose one from set $\{d_1,d_2\}$ with the best OF. For the three real solutions case, choose the solutions that fall within the range of $[d_1,d_2]$, which is denoted as $\cal S$. Then, choose the one from the set $\{d_1,d_2,\cal S\}$ with the best OF as the globally optimal solution.

Finally, the BCD method, which iterates between power allocation and location optimization, is applied to solve Problem (\ref{inDAfefro1}) for the general case. The details are omitted for simplicity.

\subsection{Special Case: $1 \ll h_ip_i, i=1,2$}

In this case, the SNR $\gamma$ can be approximated as
 \begin{equation}\label{hggtiohg}
   \gamma  \approx \frac{{{h_1}{h_2}{p_1}{p_2}}}{{{h_2}{p_2} + {h_1}{p_1}}}\buildrel \Delta \over = {\tilde \gamma }.
 \end{equation}
By substituting the expressions of ${h_1}$ and $h_2$ in (\ref{edeaftg}) into  (\ref{hggtiohg}), ${\tilde \gamma }$ can be obtained as
\begin{equation}\label{wdefr}
  \tilde \gamma  =\frac{{{\beta _1}{\beta _2}{p_1}{p_2}}}{{{\beta _2}{p_2}\left( {{H^2} + {x^2}} \right) + {\beta _1}{p_1}\left( {{H^2} + {{\left( {D - x} \right)}^2}} \right)}}.
\end{equation}
Let us denote $ x_0 = \frac{{D{\beta _1}{p_1}}}{{{\beta _1}{p_1} + {\beta _2}{p_2}}}$. The optimal $x$ that maximizes  $\tilde \gamma $ can be expressed as
\begin{equation}\label{dagtjuik}
x^* = \left\{ \begin{array}{l}
{x_0},\quad{\rm{if}}\ {d_1} \le {x_0} \le {d_2},\\
{d_1},\quad{\rm{if}}\ {x_0} \le {d_1},\\
{d_2},\quad{\rm{if}}\ {x_0} \ge {d_2}.
\end{array} \right.
\end{equation}

We consider the conditions in (\ref{dagtjuik}) case-by-case.
\subsubsection{Condition I: ${d_1} \le {x_0} \le {d_2}$}
By substituting the optimal $x^*=x_0$ into (\ref{wdefr}), Problem (\ref{inDAfefro1}) can be rewritten as
\begin{subequations}\label{convert-pro1}
\begin{align}
\mathop {\min }\limits_{\left\{ {{p_1\ge 0},{p_2\ge 0}} \right\}} \;\;\;&  \frac{{{H^2}\left( {{\beta _1}{p_1} + {\beta _2}{p_2}} \right)}}{{{p_1}{p_2}}} + \frac{{{\beta _1}{\beta _2}{D^2}}}{{{\beta _1}{p_1} + {\beta _2}{p_2}}}\\
{\rm{s.t.}}\;\;\;& D{\beta _1}{p_1} \ge {\beta _1}{d_1}{p_1} + {\beta _2}{d_1}{p_2},\label{dewdewf}\\
& D{\beta _1}{p_1} \le {\beta _1}{d_2}{p_1} + {\beta _2}{d_2}{p_2},\label{kiukoi}\\
& p_1+p_2 =  P_{T},\label{hjidhw}
\end{align}
\end{subequations}
where (\ref{hjidhw}) is due to Lemma 2.

In the following, we address Problem (\ref{convert-pro1}) by considering two cases: 1) $\beta _1=\beta _2$; 2) $\beta _1 \ne \beta _2$.

\emph{Case I: $\beta _1=\beta _2$}:
Problem (\ref{convert-pro1}) is equivalent to
\begin{subequations}\label{case1}
\begin{align}
\mathop {\max }\limits_{ {{p_1\ge 0}} } \;\;\;&  p_1(P_{T}-p_1)\\
{\rm{s.t.}}\;\;\;& \frac{{{d_2}{P_T}}}{D} \ge {p_1} \ge \frac{{{d_1}{P_T}}}{D}.
\end{align}
\end{subequations}
Obviously, the optimal solution can be obtained as follows:
\begin{equation}\label{joihuo}
p_1^* = \left\{ \begin{array}{l}
\frac{{{d_2}{P_T}}}{D},\quad{\rm{ if  }}\ {2d_2} \le D,\\
\frac{{{d_1}{P_T}}}{D},\quad{\rm{ if  }}\ {2d_1} \ge D,\\
\frac{{{P_T}}}{2},\qquad{\rm{    otherwise}}.
\end{array} \right.
\end{equation}
Then, the optimal $p_2$ is given by $p_2^*=P_T-p_1^*$.

\emph{Case II: $\beta _1\ne \beta _2$}:
The closed-form solution cannot be obtained as Case I. However, we can obtain the globally optimal solution of Problem (\ref{convert-pro1}).

\emph{\textbf{Theorem 1}}:  Problem (\ref{convert-pro1}) is a convex optimization problem.

\emph{\textbf{Proof}}: \upshape Obviously, the set of constraints in Problem (\ref{convert-pro1}) is linear. Hence, we only need to prove the convexity of the OF of Problem (\ref{convert-pro1}).

Denote OF of Problem (\ref{convert-pro1}) as function $f(p_1,p_2)$. Obviously, $f(p_1,p_2)$ is twice differentiable, and its Hessian matrix can be derived as
\begin{equation}\label{ewfrfre}
\!\!{\nabla ^2}f \!\!=\!\! \left[ {\begin{array}{*{20}{c}}
{\!\!\frac{{2{H^2}{\beta _2}}}{{p_1^3}} + \frac{{2\beta _1^3{\beta _2}{D^2}}}{{{{\left( {{\beta _1}{p_1} + {\beta _2}{p_2}} \right)}^3}}}}&{\frac{{2\beta _1^2\beta _2^2{D^2}}}{{{{\left( {{\beta _1}{p_1} + {\beta _2}{p_2}} \right)}^3}}}}\\
{\frac{{2\beta _1^2\beta _2^2{D^2}}}{{{{\left( {{\beta _1}{p_1} + {\beta _2}{p_2}} \right)}^3}}}}&{\frac{{2{H^2}{\beta _1}}}{{p_2^3}} + \frac{{2{\beta _1}\beta _2^3{D^2}}}{{{{\left( {{\beta _1}{p_1} + {\beta _2}{p_2}} \right)}^3}}}\!\!\!\!}
\end{array}} \right]
\end{equation}
and its determinant is checked to be strictly bigger than zero. In addition,  both the diagonal elements are strictly positive. Hence, ${\nabla ^2}f$ is positive definite. The proof  completes. \hfill\rule{2.7mm}{2.7mm}

The globally optimal solution can be obtained by using standard convex optimization algorithms such as interior-point method \cite{boyd2004convex}.

\vspace{-0.1cm}
\subsubsection{Condition II: ${x_0}<{d_1}$} By substituting $x^*=d_1$ into (\ref{wdefr}) and using Lemma 2, Problem (\ref{inDAfefro1}) can be rewritten as
\begin{subequations}\label{CEREFTRooi}
\begin{align}
\mathop {\max }\limits_{p_1} \;\;\;&  \frac{{{\beta _1}{\beta _2}{p_1}({P_T} - {p_1})}}{{({\beta _1}{D_2} - {\beta _2}{D_1}){p_1} + {\beta _2}{D_1}{P_T}}}\\
{\rm{s.t.}}\;\;\;& 0 \le {p_1} \le p_1^{{\rm{up}}},\label{hsadwafrhw}
\end{align}
\end{subequations}
where ${D_1} = {H^2} + {d_1^2}$, ${D_2} = {H^2} + {\left( {D -  d_1} \right)^2}$, and $p_1^{{\rm{up}}} = \frac{{{d_1}{\beta _2}{P_T}}}{{\left( {D - {d_1}} \right){\beta _1} + {d_1}{\beta _2}}}$.

We solve this problem by considering two cases: 1) ${\beta _1}{D_2} = {\beta _2}{D_1}$; 2) ${\beta _1}{D_2}  \ne {\beta _2}{D_1}$.

\emph{Case I:${\beta _1}{D_2} = {\beta _2}{D_1}$}: The optimal solution of Problem (\ref{CEREFTRooi}) can be obtained as follows:
\begin{equation}\label{dWDEFR}
 p_1^* = \left\{ \begin{array}{l}
\frac{{{P_T}}}{2},\ {\rm{if }}\ p_1^{{\rm{up}}} > \frac{{{P_T}}}{2}\\
p_1^{{\rm{up}}},\ {\rm{otherwise}}.
\end{array} \right.
\end{equation}

\emph{Case II: ${\beta _1}{D_2}  \ne {\beta _2}{D_1}$}: The OF of Problem (\ref{CEREFTRooi}) can be rewritten as:
\begin{equation*}
 \tilde \gamma  = \frac{{{\beta _1}{\beta _2}}}{{{\beta _1}{D_2} \!-\! {\beta _2}{D_1}}}\left( { \!-\! \left( {{p_1}\! +\! E} \right) \!-\! \frac{{E\left( {E \!+\! {P_T}} \right)}}{{{p_1} \!+\! E}} \!+\! 2E \!+\!{P_T}} \right),
\end{equation*}
where $E$ is equal to $ \frac{{{\beta _2}{D_1}{P_T}}}{{{\beta _1}{D_2} - {\beta _2}{D_1}}}$. The second derivative of $ \tilde \gamma$ w.r.t. $p_1$ is given by
\begin{equation}\label{WEDfoilio}
\frac{{{\partial ^2}\tilde \gamma }}{{\partial p_1^2}} =  - \frac{{2{\beta _1}{\beta _2}}}{{{\beta _1}{D_2} - {\beta _2}{D_1}}}\frac{{E\left( {E + {P_T}} \right)}}{{{{\left( {{p_1} + E} \right)}^3}}}
\end{equation}
which is proved to be negative. Hence, Problem (\ref{CEREFTRooi}) is a convex optimization problem. Define ${{\bar p}_1} \triangleq \sqrt {E\left( {E + {P_T}} \right)}  - E$. The optimal solution is given by
\begin{equation}\label{dsacR}
 p_1^* = \left\{ \begin{array}{l}
{{\bar p}_1},\ {\rm{if }}\ p_1^{{\rm{up}}} > {{\bar p}_1}\\
p_1^{{\rm{up}}},\ {\rm{otherwise}}.
\end{array} \right.
\end{equation}

\vspace{-0.2cm}
\subsubsection{Condition III: ${x_0}>{d_2}$}
The optimal solution in this case can be obtained by using the similar method as those in Condition II, the details of which are omitted here.

When the optimal solution for each condition is obtained, select one solution with the largest value of ${\tilde \gamma }$ as the globally optimal solution of Problem (\ref{inDAfefro1}).

\section{3-D Channel Model}\label{jojt}

In this section, we extend the free-space channel model to 3-D channel model proposed in \cite{al2014optimal}, where the impacts of blockage and shadowing are taken into account and is more practical than free-space channel model. In specific, the line-of-sight (LoS) probability is given by
\begin{equation}\label{dwerji}
\vspace{-0.2cm}
  {P_{{\rm{LoS}}}} = \frac{1}{{1 + a\exp \left( { - b\left( {\theta  - a} \right)} \right)}},
\end{equation}
where $a$ and $b$ are positive environment-related parameters and $\theta$ is the elevation angle between the UAV and the ground devices (controller or robot) as shown in Fig.~\ref{systemodel}. Some typical values of $a$ and $b$ can be found in Table I of \cite{bor2016efficient}. It is observed from (\ref{dwerji}) that the LoS probability increases with the elevation angle, which is reasonable as the probability that signal is blocked is decreasing when the height of UAV is increasing.

When the location of one UAV is given, the mean path loss is given by \cite{al2014optimal}:
\vspace{-0.1cm}
\begin{equation}\label{xswdff}
\vspace{-0.1cm}
  L(\theta ,d) = \frac{A}{{1 + a\exp \left( { - b\left( {\theta  - a} \right)} \right)}} + 20{\log _{10}}\left( {{d}} \right) + C,
\end{equation}
where $A$ and $C$ are constants given by $A=\eta _{{\rm{LoS}}}-\eta _{{\rm{NLoS}}}$ and $C=20{\log _{10}}\left( {\frac{{4\pi {f_c}}}{c}} \right) + {\eta _{{\rm{NLoS}}}}$, respectively. $d$ is the distance between the UAV and the ground devices (controller or robot). $\eta _{{\rm{LoS}}}$ and $\eta _{\rm{NLoS}}$ are the path loss (in dB) corresponding to the LoS and non-LoS (NLoS) links. In general, $\eta _{\rm{NLoS}}$ is larger than $\eta _{\rm{LoS}}$ due to the more severe attenuation associated with NLoS. $f_c$ is the central frequency point, $c$ is the light speed.

Based on the path loss model in (\ref{xswdff}), the normalized channel gains w.r.t. noise power are given by
\begin{equation}\label{jjgjngkk}
  h_i=\tilde C_i{d_i^{ - 2}}{10^{  \frac{{\tilde A_i}}{{1 + a_i\exp \left( { - b_i(\theta_i  - a_i)} \right)}}}}, i=1,2
\end{equation}
where  $\tilde A_i = -\frac{{A_i}}{{10}}>0$ and $\tilde C_i ={{{{10}^{ - \frac{C_i}{{10}}}}} \mathord{\left/
 {\vphantom {{{{10}^{ - \frac{C_i}{{10}}}}} {{\delta ^2}}}} \right.
 \kern-\nulldelimiterspace} {{\delta ^2}}}$ with $\delta ^2$ denoting the noise power, and $\theta_i$ are given by
\begin{equation}\label{hyuik}
 {\theta _1} = \arctan \left( {\frac{H}{x}} \right),{\theta _2} = \arctan \left( {\frac{H}{{D - x}}} \right).
\end{equation}

Similar to the free-space case, we also adopt the BCD algorithm to solve Problem (\ref{inDAfefro1}). When $x$ and $H$ are given, channel gains $h_1$ and $h_2$ are fixed. Then, the power allocation can be optimized by using the same method in Subsection \ref{joijiy}. In the following, we only focus on the optimization of height $H$ and horizontal distance $x$ when the other parameters are fixed.

\subsection{Optimization of $H$ with fixed $x$, $p_1$ and $p_2$}
\begin{figure}
\centering
\includegraphics[width=2.4in]{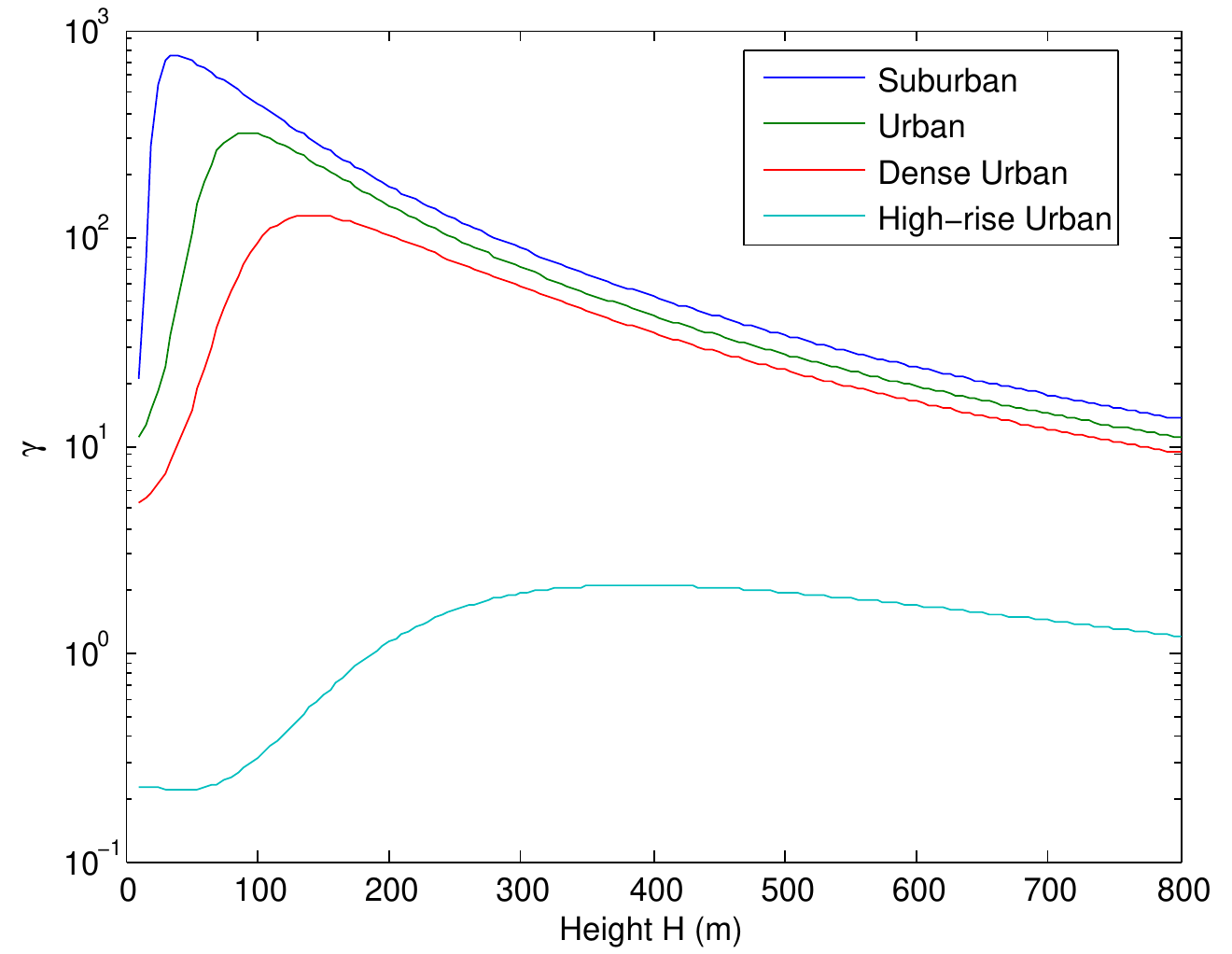}
\caption{SINR $\gamma$ versus height $H$ when $x=100$ m.}
\vspace{-0.3cm}
\label{checkH}
\end{figure}

When $x$, $p_1$ and $p_2$ are given, the SNR $\gamma(H)$ is a very complicated function of $H$. It is difficult to strictly prove the monotonically and convexity of this function. As in \cite{bor2016efficient} and \cite{cpan2019}, we graphically illustrate these properties in Fig.~\ref{checkH}, where we show the SINR $\gamma(H)$ versus $H$ with $x=100$ m. Four different scenarios are illustrated, and the corresponding parameters for each scenario are given in \cite{bor2016efficient}. It can be found from this figure that for each scenario, the SINR value first increases with height $H$ and then decreases with $H$. As a result, there exists only one maximum point for each scenario, denoted as $H^\star$.  The value of $H^\star$ is the solution to the following equation:
\begin{equation}\label{hjik}
  \frac{{d\gamma (H)}}{{dH}} = 0.
\end{equation}
Similar to \cite{bor2016efficient}, the bisection search method can be used to find the root of the above equation.
\subsection{Optimization of $x$ with fixed $H$, $p_1$ and $p_2$}
\begin{figure}
\centering
\includegraphics[width=2.4in]{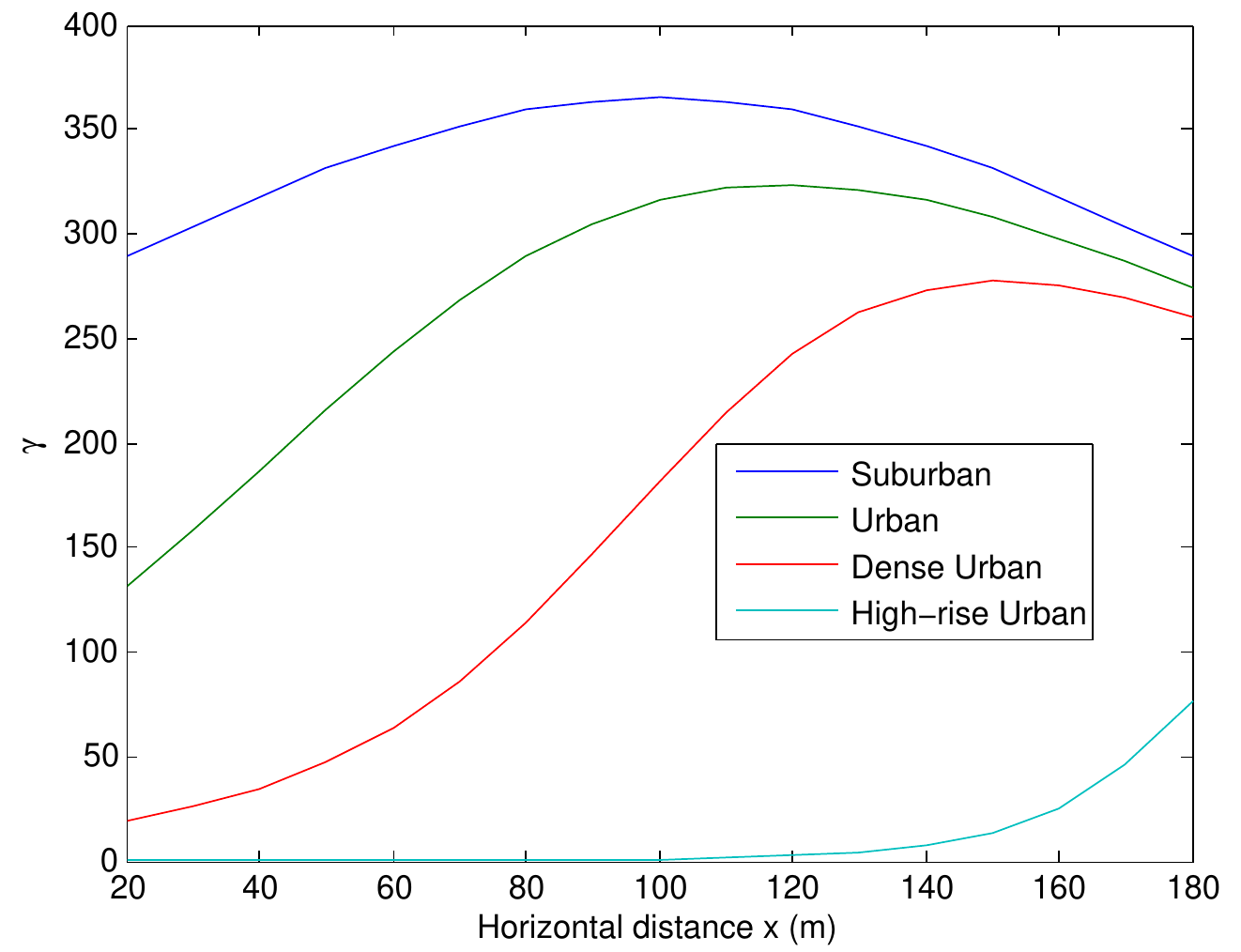}
\caption{SINR $\gamma$ versus the horizontal distance $x$ when $H=120$ m.}
\vspace{-0.3cm}
\label{checkx}
\end{figure}

In Fig.~\ref{checkx}, we illustrate the SINR value versus the horizontal distance $x$ when $H=120$ m. The channel from the controller the UAV is assumed to be suburban environment. The SNR values when the channel  from the UAV to the robot experiences various environments are shown in Fig.~\ref{checkx}. Similar to Fig.~\ref{checkH}, the SINR value also first increases with $x$ and then decreases with $x$, or always increases with $x$. Then, the bisection search method can be adopted to find the optimal solution.
\section{Simulation Results}\label{simlresult}
Simulation results are performed to check the performance of the proposed algorithms. The system parameters are set as  $D=200$ m, $H=120$ m, $d_1=30$ m, $d_2=170$ m, $L=100$ bits,  $M=80$,  and $P_T=4$ Watt.

\subsection{Free-space Channel Model}
We first study the free-space channel model, where $\beta_1=50$ dB, $\beta_2=59$ dB.
\begin{figure}
\centering
\includegraphics[width=2.4in]{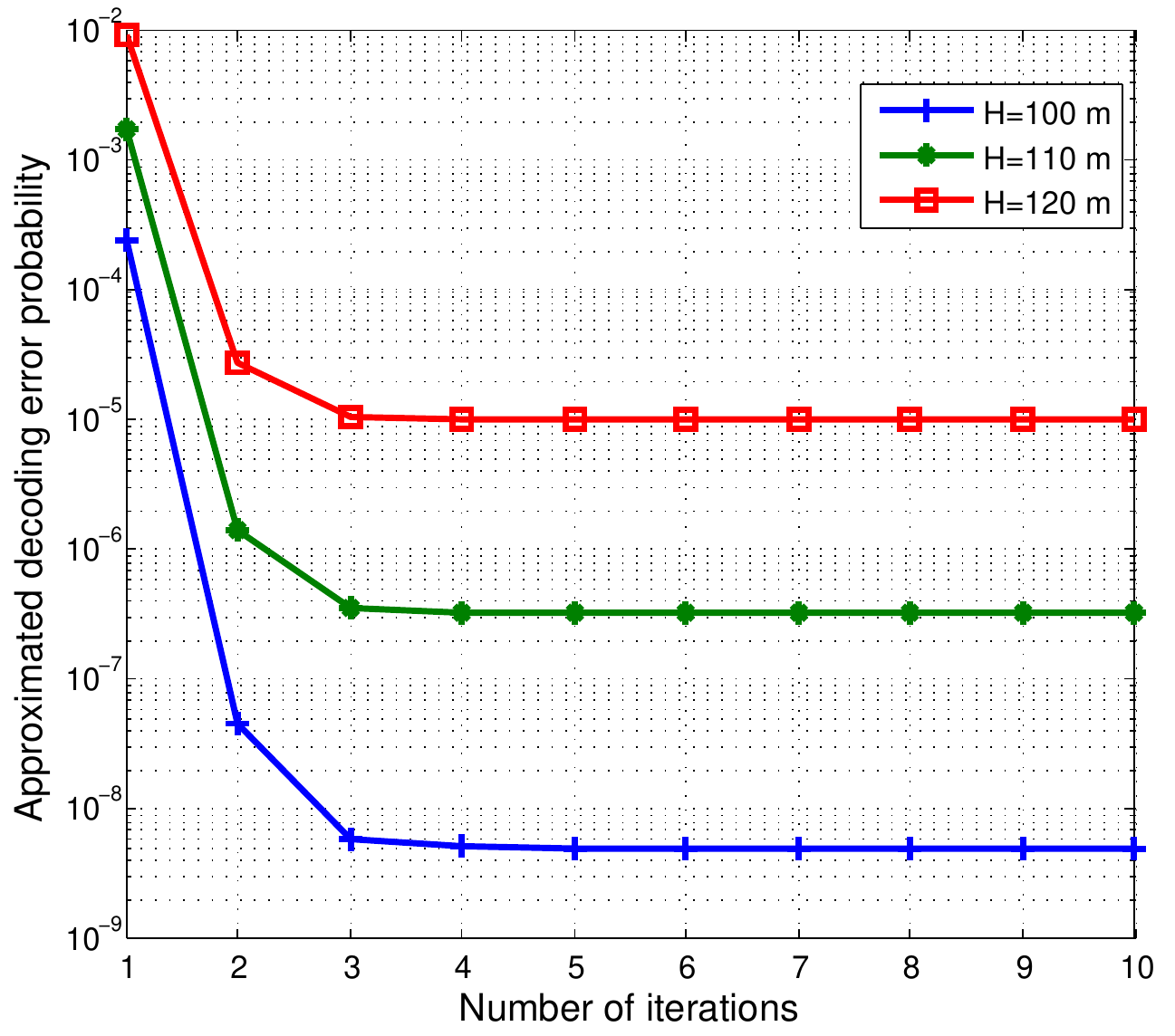}
\caption{Convergence behaviour.}
\vspace{-0.4cm}
\label{convergence}
\end{figure}

In Fig.~\ref{convergence}, we study the convergence behaviour of the iterative algorithm for the general case. It is shown that the proposed iterative algorithm converges rapidly and in general four iterations are sufficient for the algorithm to converge, which implies  low complexity of our proposed algorithm.

\begin{figure}
\centering
\includegraphics[width=2.4in]{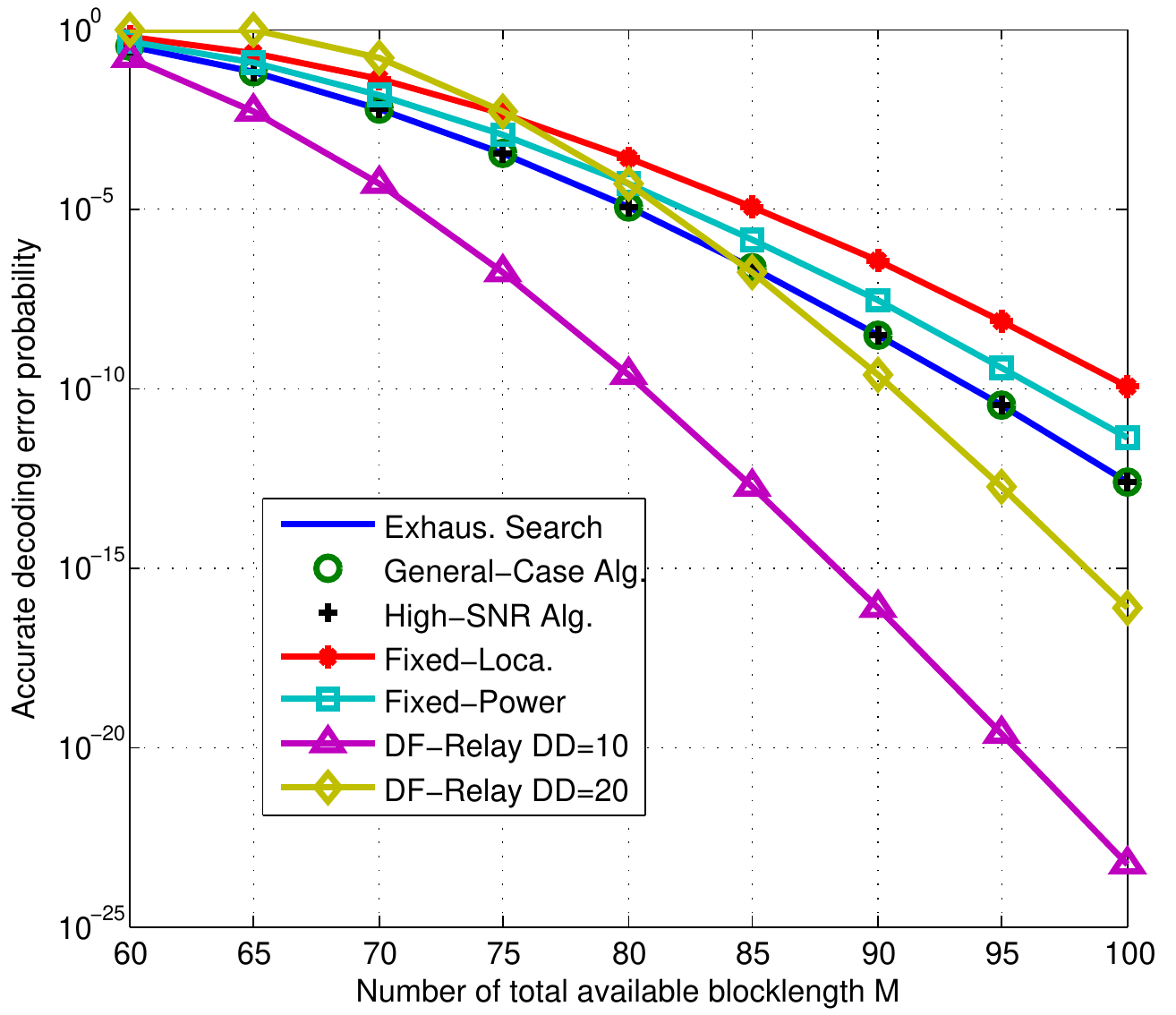}
\caption{Performance comparison, $H=120$ m.}
\vspace{-0.4cm}
\label{Perforcompare}
\end{figure}

In Fig.~\ref{Perforcompare}, we compare the performance of various algorithms, which include: 1) General case (`General-Case Alg.'); 2) High-SNR case (`High-SNR Alg.'); 3) Exhaustive search algorithm (`Exhaus. Search'); 4) Fixed location with $x=(d_1+d_2)/2$ (`Fixed-Loca.'); 5) Fixed power allocation with $p_1=p_2=P_T/2$ (`Fixed-Power'). We also compare the proposed algorithm for AF relay with the one for DF relay in \cite{cpan2019}. For the DF relay, the number of channel uses for the signal processing at the relay is denoted as DD \footnote{In general, the system bandwidth is fixed, and then the number of channel users can be interpreted as time duration.}. It is observed in Fig.~\ref{Perforcompare} that the proposed two algorithms  significantly outperform  the Fixed-Loca. algorithm and Fixed-Power algorithm, which confirms the benefits of our proposed algorithms. It is interesting to find that the proposed two algorithms have almost the same performance as the exhaustive search method. This may be due to the fact that the SNR  in this example generally operates in a very high regime. When the signal processing delay DD is small (e.g. DD=10), the DF relay outperforms the AF relay, which means DF relay is a good option. On the other hand, when DD is large, the AF relay performs better than the DF relay when the number of channel blocklength $M$ is small. In this example, when DD=20, the performance of the AF relay is better than that of the DF relay when $M\le 85$. This means that it is beneficial to adopt the AF relay when the latency requirement is stringent, which is usually the case in URLLC applications. The reason is that when more time is used for signal processing, the left time for data transmission will be reduced, which decreases the reliability performance.

\subsection{3-D channel model}
\begin{figure}
\centering
\includegraphics[width=2.4in]{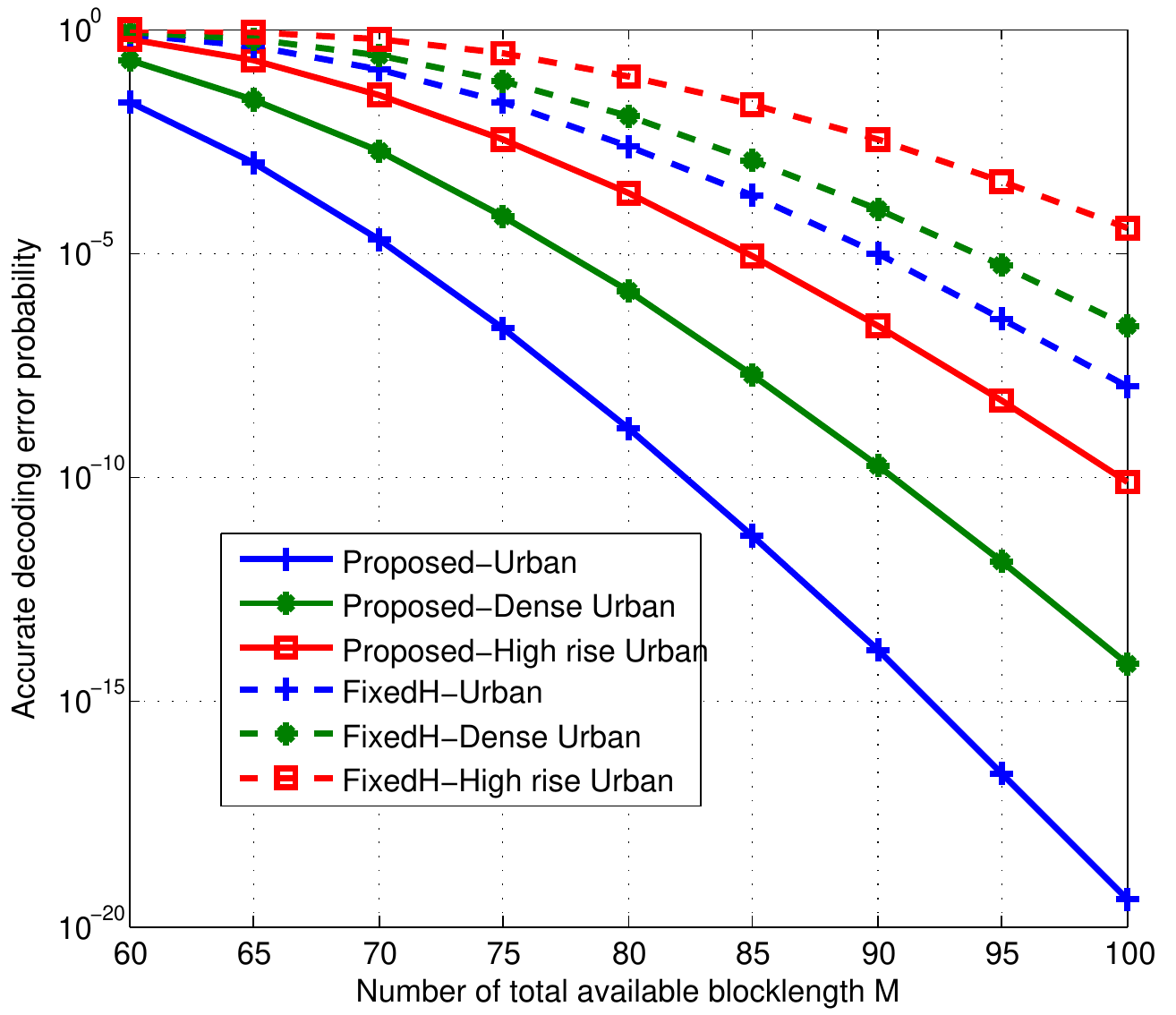}
\caption{Performance comparison for 3-D channel models.}
\vspace{-0.4cm}
\label{Perforcompare3D}
\end{figure}

In Fig.~\ref{Perforcompare3D}, we study the performance of the algorithm developed in Section \ref{jojt} for the 3-D channel model.
The simulation parameters are set as follows: $H_{\min}=10$ m, $H_{\max}=200$ m, $d_1=20$ m, $d_2=200$ m, $f_c=2.5$ GHz, noise power is -93 dB. The channel model from the controller the UAV is assumed to be suburban environment. We study the performance when the channel  from the UAV to the robot experiences various environments. The environment parameters are referred to \cite{bor2016efficient}. To study the importance of optimizing the height, we also show the performance when the height is fixed as $H=100$ m, which is denoted as `FixedH' in Fig.~\ref{Perforcompare3D}. It can be observed that the proposed joint optimization algorithm outperforms the the `FixedH' algorithm for various environments, and the performance gains increase with $M$, which confirms the importance of optimizing the height.

\vspace{-0.1cm}
\section{Conclusions}\label{conclu}
\vspace{-0.1cm}
This paper studied the joint power allocation and location optimization for UAV AF-relay system with URLLC requirements. Both the free-space channel and the 3-D channel are considered. For the free-space channel, the iterative algorithm was proposed for general case, and the closed-form solution was derived for the special case with high SNR. A low-complexity iterative algorithm was proposed for the 3-D channel model. Simulation results showed that the proposed algorithms can achieve the same performance as the exhaustive search method, and outperform the other algorithms such as fixed location or fixed power allocation.
\vspace{-0.2cm}
\numberwithin{equation}{section}
\begin{appendices}
\section{Proof of Lemma 1}\label{prooflemma1}
The first derivative of $\varepsilon$ w.r.t. $\gamma$ is
\begin{equation}\label{daef}
\varepsilon'=  - \frac{1}{{\sqrt {2\pi } }}{e^{ - \frac{{f^2({\gamma})}}{2}}}f'(\gamma),
\end{equation}
where ${f}({\gamma })$ is short for function $f\left( {\gamma, m, L} \right)$ and $f'(\gamma)$ is the first derivative of ${f}({\gamma })$ w.r.t. $\gamma$ that is given by
\begin{eqnarray}
  \!\!\! \!\!\!\!\!\! f'(\gamma) \!\!\! &\!= \!&\frac{{\sqrt m }}{{\sqrt {{{(1 + \gamma )}^2} - 1} }}\left( {1\! -\! \ln 2\frac{{{{\log }_2}(1 \!+\! \gamma )\! -\! \frac{L}{m}}}{{{{(1 + \gamma )}^2} - 1}}} \right) \label{deafrg}\\
   &\ge& \frac{{\sqrt m }}{{\sqrt {{{(1 + \gamma )}^2} - 1} }}\left( {1 - \frac{{{{\ln }}(1 + \gamma )}}{{{{(1 + \gamma )}^2} - 1}}} \right).\label{jytijju}
\end{eqnarray}
Let $x=1+\gamma$ and thus $x\ge1$. Define function $g(x)$ as
\begin{equation}\label{dwfrfr}
  g(x)\buildrel \Delta \over = \frac{{{{\ln }}(x )}}{{{{x }^2} - 1}}.
\end{equation}
The first derivative of $g(x)$ w.r.t. $x$ is given by
\begin{equation}\label{defrfregt}
 g'(x) = \frac{{G(x)}}{{x{{({x^2} - 1)}^2}}}
\end{equation}
where $G(x) = {x^2} - 1 - 2{x^2}\ln x$. The first derivation of $G(x)$ w.r.t. $x$ is given by $G'(x)=-4x\ln(x)\le 0$ for $x\ge 1$. Hence, $G(x)$ is a decreasing function for $x\ge 1$ and thus $G(x)\le G(1)=0$ holds. Please note that the denominator of (\ref{defrfregt}) is positive, then $g'(x)\le 0$ holds for $x\ge 1$. Hence $g(x)$ is a decreasing function of $x$ and $g(x)\le g(1)$. By using the L'Hospital's rule, $g(1)$ can be calculated as $g(1)=1/2$. By plugging the inequality $g(x)\le 1/2$ into (\ref{jytijju}), we obtain
\begin{equation}\label{dwfrrf}
  f'(\gamma)\ge \frac{{\sqrt m }}{2{\sqrt {{{(1 + \gamma )}^2} - 1} }}\ge 0.
\end{equation}
Hence, $\varepsilon'\le 0$, which completes the proof.

%
\end{appendices}


\
\






\vspace{-0.3cm}
\bibliographystyle{IEEEtran}
\vspace{-0.2cm}
\bibliography{myre}


\end{document}